\documentclass[%
 reprint,
 amsmath,amssymb,
 aps,prl
]{revtex4-2}

\usepackage{graphicx}
\usepackage{tikz}
\usetikzlibrary{decorations.pathmorphing}
\usepackage{dcolumn}
\usepackage{bm}
\usepackage{physics}
\usepackage{simplewick}
\usepackage[dvipsnames]{xcolor}
\usepackage{braket,cancel}
\definecolor{lgray}{gray}{0.4}
\definecolor{cor}{rgb}{0.94, 0.41, 0.35}
\definecolor{mag}{rgb}{0.7451, 0.2039, 0.3333}
\usepackage[colorlinks=true,urlcolor=mag,linkcolor=mag,
citecolor=mag,pdfpagelabels=true,hypertexnames=true,
plainpages=false,naturalnames=false]{hyperref}
\newcommand{\x}{{\boldsymbol x}}

\def\k{{\boldsymbol k}}

\def\be{\begin{equation}}
\def\ee{\end{equation}}
\def\ba{\begin{array}}
\def\ea{\end{array}}
\def\nn{\nonumber}

\begin{document}

\title{Pushing the Primordial Frontier: Exact Linear Solutions in Multifield Inflation}

\author{Javier Huenupi$^a$}

\author{Claudio Mu\~noz$^b$}

\author{Gonzalo A. Palma$^c$}

\author{Spyros Sypsas$^{d}$}

\affiliation{
$^a$ Center for Particle Cosmology, Department of Physics and Astronomy, University of Pennsylvania, Philadelphia, PA 19104, USA
\\
$^b$ Departamento de Ingenier\'ia Matem\'atica and Centro de Modelamiento Matem\'atico (IRL
2807 CNRS), Universidad de Chile, Casilla 170 Correo 3, Santiago, Chile
\\
$^c$ Departamento de F\'isica, FCFM, Universidad de Chile, Blanco Encalada 2008, Santiago, Chile
\\
$^d$Departamento de Ciencias B\'asicas and Centro de Ciencias Exactas, Facultad de Ciencias, Universidad del B\'io-B\'io, Chill\'an, Chile
 }

\begin{abstract}  
We present exact analytic solutions for the linear dynamics of a two-field inflationary system in which the primordial curvature perturbation $\zeta$ is coupled to an isocurvature perturbation $\sigma$ of entropy mass $\mu$. The solutions are valid for arbitrary values of $\mu$ and the dimensionless interaction strength $\lambda$. They therefore provide analytic control over the strong mixing regime in which $\zeta$ interacts with light isocurvature fields, commonly associated with rapid-turn inflation. As a first application, we derive the amplitude of the primordial power spectrum in closed form, obtaining an expression that interpolates between the weakly mixed, strongly mixed, light-field, and heavy-field regimes. These results open the way to analytic studies of multifield observables beyond the power spectrum, including non-Gaussianity, particle production, and loop corrections.
\end{abstract}

%\keywords{Primordial cosmology, cosmic inflation, multifield inflation, rapid-turn inflation, cosmological colliders}
                          
\maketitle

\noindent \emph{\bf Introduction.--}
It is hard to imagine that the primordial curvature perturbation $\zeta$---the fluctuation responsible for seeding the large-scale structure of the universe---evolved in complete isolation during inflation~\cite{Guth:1980zm, Starobinsky:1980te, Linde:1981mu, Albrecht:1982wi, Achucarro:2022qrl}. Multifield inflation provides the natural framework for describing interactions between $\zeta$ and other degrees of freedom, as encountered in supergravity and string-theory descriptions of the primordial universe~\cite{Binetruy:1986ss, Brustein:1992nk, Banks:1995dp, Kachru:2003sx, Blanco-Pillado:2004aap, Lalak:2005hr, Dimopoulos:2005ac, Conlon:2005jm, Bond:2006nc, Grimm:2007hs, Covi:2008cn, Ellis:2013xoa, Carrasco:2015rva, Hetz:2016ics, Aragam:2021scu}.  How strong were these interactions, and what imprints could they have left on cosmological observables? A vast body of literature has addressed these questions, uncovering a wide range of phenomena, including features in the primordial spectra~\cite{Achucarro:2010da, Gao:2012uq, Achucarro:2012fd, Gao:2013ota, Noumi:2013cfa, Achucarro:2013cva, Mizuno:2014tza, Achucarro:2014msa, Gao:2015aba, Boutivas:2022qtl}, enhanced primordial non-Gaussianity~\cite{Seery:2005gb, Rigopoulos:2005ae, Vernizzi:2006ve, Byrnes:2008wi, GarciaSaenz:2018vqf, Chen:2018brw, Fumagalli:2019noh, Bjorkmo:2019aev, Garcia-Saenz:2019njm, Achucarro:2021pdh, Wang:2022eop, Iarygina:2023rapid, Aoki:2026qea}, particle production~\cite{Lalak:2007vi, Shiu:2011qw, Konieczka:2014zja, Achucarro:2016fby, Achucarro:2019mea, Achucarro:2019lgo, Fumagalli:2020nvq, Firouzjahi:2021lov, Fumagalli:2021mpc, Parra:2024usv, Achucarro:2026qyx}, and mechanisms for generating primordial black holes~\cite{Garcia-Bellido:1996mdl, Palma:2020ejf, Fumagalli:2020adf, Braglia:2020eai, Anguelova:2020nzl, Geller:2022nkr, Bhattacharya:2022fze, Lorenzoni:2025gni}. Although multifield inflation plays a central role in exploring the early universe beyond the single-field paradigm, analytic progress has been limited by the absence of exact curvature and isocurvature mode functions~\cite{Sugiyama:2011jt, Kaiser:2013sna, Brown:2017osf, Mizuno:2017idt, Achucarro:2018ngj, Nguyen:2019kbm}. Interactions among the different degrees of freedom lead to a nontrivial system of equations~\cite{Gordon:2000hv, GrootNibbelink:2001qt, Langlois:2008qf} that has restricted most analytic treatments to regimes where perturbative methods or effective single-field descriptions apply~\cite{Chen:2009zp, Tolley:2009fg, Chen:2009we, Cremonini:2010ua, Baumann:2011nk, Cespedes:2012hu, Achucarro:2012sm, Chen:2012ge, Achucarro:2012yr, Noumi:2012vr, Castillo:2013sfa, Gwyn:2012mw, Iyer:2017qzw,An:2017hlx}. The phenomenologically rich regime wherein $\zeta$ interacts strongly with light isocurvature fields has therefore remained comparatively unexplored analytically, despite extensive numerical and approximate studies~\cite{RenauxPetel:2015mga, GarciaSaenz:2018ifx, Aragam:2019omo, Chakraborty:2019dfh, Bjorkmo:2019fls, Aragam:2020uqi, Christodoulidis:2023eiw, Wolters:2024rapid, Anguelova:2024akm}.

In this Letter, we present exact analytic solutions for the linear dynamics of a two-field inflationary system in which the primordial curvature perturbation $\zeta$ is coupled, with dimensionless strength $\lambda$, to an isocurvature perturbation $\sigma$ of entropy mass $\mu$. To our knowledge, these are the first closed-form solutions for an inflationary system of this type that remain valid for arbitrary values of $\lambda$ and $\mu$. As a first application, we derive the dimensionless primordial power spectrum $\Delta_\zeta$ in closed form.

\begin{figure}[t]
\centering
\includegraphics[width=\columnwidth]{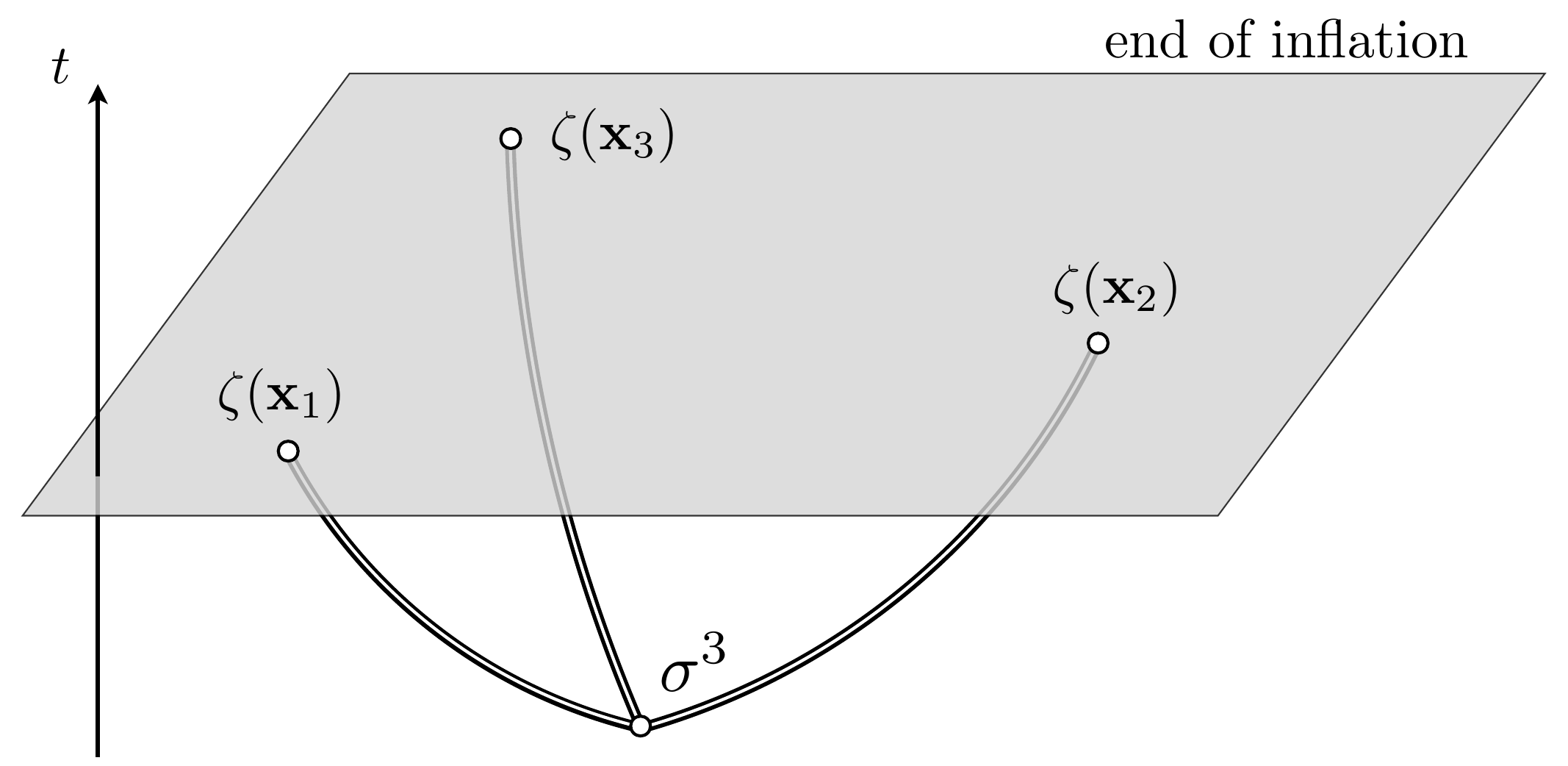}
\caption{Three exact mixed propagators connect an illustrative cubic $\sigma$ vertex to late-time curvature perturbations.}
\label{fig:quadratic-resummation}
\vspace{-6pt}
\end{figure}

Viewed through the cosmological-collider lens, these solutions have a simple interpretation. Extra particles present during inflation can leave traces in $n$-point correlators of primordial fluctuations that can be constrained by observations~\cite{Planck:2013jfk, Planck:2013wtn}. Calculating these traces requires following how these particles mix with the curvature perturbation. Our exact solutions sum the infinite series of quadratic interaction insertions into a single exact propagator, whose mixed $\sigma$--$\zeta$ component, schematically, is:
\begin{equation}
\label{eq:quadratic-resummation}
\begin{tikzpicture}[x=1cm,y=1cm,baseline=-0.5ex,
  every node/.style={inner sep=1pt},
  sigma/.style={decorate,decoration={snake,amplitude=1pt,segment length=4pt},line width=0.5pt},
  zeta/.style={line width=0.5pt}]
\draw[line width=0.55pt,double distance=0.8pt] (0,0)--(0.95,0);
\node[anchor=east,inner sep=1pt] at (-0.05,0) {$\sigma$};
\node[anchor=west,inner sep=1pt] at (1.05,0) {$\zeta$};
\filldraw[fill=white,draw=black,line width=0.45pt] (0,0) circle (1.6pt);
\filldraw[fill=white,draw=black,line width=0.45pt] (0.95,0) circle (1.6pt);
\begin{scope}[xshift=0.25cm]
\node at (1.25,0) {$=$}; 
\draw[sigma] (1.55,0)--(2.15,0);
\draw[zeta] (2.15,0)--(2.75,0);
\fill (2.15,0) circle (1.3pt);
\node[above=3pt,font=\scriptsize] at (2.15,0) {$\lambda$};
\node at (3.05,0) {$+$};
\draw[sigma] (3.35,0)--(3.95,0);
\draw[zeta] (3.95,0)--(4.55,0);
\draw[sigma] (4.55,0)--(5.15,0);
\draw[zeta] (5.15,0)--(5.75,0);
\foreach \pos in {3.95,4.55,5.15}{
  \fill (\pos,0) circle (1.3pt);
  \node[above=3pt,font=\scriptsize] at (\pos,0) {$\lambda$};}
\node at (1.25,-0.62) {$+$};
\draw[sigma] (1.55,-0.62)--(2.15,-0.62);
\draw[zeta] (2.15,-0.62)--(2.75,-0.62);
\draw[sigma] (2.75,-0.62)--(3.35,-0.62);
\draw[zeta] (3.35,-0.62)--(3.95,-0.62);
\draw[sigma] (3.95,-0.62)--(4.55,-0.62);
\draw[zeta] (4.55,-0.62)--(5.15,-0.62);
\foreach \pos in {2.15,2.75,3.35,3.95,4.55}{
  \fill (\pos,-0.62) circle (1.3pt);
  \node[above=3pt,font=\scriptsize] at (\pos,-0.62) {$\lambda$};}
\node at (5.45,-0.62) {$+$};
\node at (5.85,-0.62) {$\cdots$};
\end{scope} 
\end{tikzpicture} 
\end{equation}
where the double line denotes the exact mixed propagator, whereas the wavy and solid lines denote free $\sigma$ and $\zeta$ propagators. Interactions of cubic and higher order can then be computed using this propagator, without expanding in the mixing strength (Fig.~\ref{fig:quadratic-resummation}). This gives analytic access to particle signatures even when mixing cannot be treated as a small correction, as illustrated in the companion article~\cite{Huenupi:2026aqc}, which develops Schwinger-Keldysh diagrammatic rules valid to all orders in $\lambda$, and uses them to derive tree-level primordial bispectra.

%%%%%%%%%%%%%%%%%%%%%%%%%%%%%%%%%%%%%%%%%%%%%%%%%%%%%%%%%%%%%%%%%%%%%%%%%%%%%%%%%%%%%%%%%%%%%%%%%%%%%%%%%%%%%%%%%%%%%%%%%%%%%%%%%%%%%%%%%%%%%%%%%%%%%%%%%%%%%%%%%%%%%%%%%%%%%%%%%%%%%%%%%%%%%%%%%%%%%%%%%%%%%%%%%%%%%%%%%%%%%%%%%%%%%%%%%%%%%%%%%%%%%%%%%%%%%%%%%%%%%%%%%%%%%%

\noindent \emph{\bf Basic Setup.--}
In the simplest two-field realization of multifield inflation, the scalar sector contains the primordial curvature perturbation $\zeta$ and a single isocurvature perturbation $\sigma$ of entropy mass $\mu$. We work in units where $c=\hbar=M_{\rm Pl}=1$. The quadratic action governing the linear dynamics of this system is~\cite{Achucarro:2016fby}
\begin{align}
S = \int  \dd[4]{x}  a^3 \bigg[ & \epsilon \left( \dot \zeta - \lambda \frac{H}{\sqrt{2 \epsilon}} \sigma \right)^2 - \frac{\epsilon}{a^2} (\nabla \zeta)^2 \nn \\ 
&
+ \frac{1}{2} \dot \sigma^2 - \frac{1}{2 a^2} (\nabla \sigma)^2 - \frac{1}{2} \mu^2 \sigma^2 \bigg] , \label{action}
\end{align}
where a dot denotes a derivative with respect to cosmic time $t$. Here $a=a(t)$ is the scale factor, $H=\dot a/a$ is the Hubble expansion rate, $\epsilon=-\dot H/H^2\ll 1$ is the first slow-roll parameter, and $\lambda$ is the dimensionless coupling controlling the interaction between $\zeta$ and $\sigma$. The action \eqref{action} describes fluctuations around a homogeneous background trajectory in a nonlinear sigma-model target space. In this setting, $\lambda$ is related to the turning rate (or more precisely, the rate of geodesic deviation) $\Omega$ of the inflationary trajectory through $\lambda = - 2 \Omega/ H$. The field-space curvature enters only through $\mu$~\cite{note-mass}. We assume a background in which $\epsilon$, $\lambda$ and $\mu$ vary slowly. 

To quantize the system, it is convenient to introduce the canonically normalized curvature perturbation
\be
\varphi \equiv \sqrt{2 \epsilon} \zeta .
\ee
Neglecting slow-roll suppressed corrections from the time dependence of $\epsilon$, the canonical momenta conjugate to $\varphi$ and $\sigma$ are $\Pi_\varphi  = a^3 ( \dot \varphi - \lambda H \sigma)$ and $\Pi_\sigma = a^3 \dot \sigma$. The fields and their canonical momenta obey the standard equal-time commutation relations of QFT. These relations are implemented by expanding the fields as
\begin{align}
\varphi (\x, t) \! &= \!\! \sum_{b} \! \int_{\bf k} \!\! \bigg[  \varphi_{b}(k, t) \hat a_b (\k) \! + \! \varphi^*_{b}(k, t) \hat a^\dag_b (-\k) \bigg] e^{i {\bf k} \cdot {\bf x}},  \label{quant-1} \\
\sigma (\x , t) \! &=  \!\! \sum_{b} \! \int_{\bf k} \!\! \bigg[ \sigma_{b}(k , t) \hat a_b (\k) \! + \! \sigma^*_{b}(k , t) \hat a^\dag_b (-\k) \bigg] e^{i {\bf k} \cdot {\bf x}},  \label{quant-2}
\end{align}
where $\int_{\bf k}\equiv(2 \pi)^{-3}\int \dd[3]{k}$. The index $b=1,2$ labels the two independent mode solutions, and $\varphi_b(k,t)$ and $\sigma_b(k,t)$ are the corresponding mode functions. On the other hand, the creation and annihilation operators satisfy $\big[ \hat a_b (\k)   , \hat a_c^{\dag} (\k') \big] = \delta_{bc} (2 \pi)^3 \delta^{(3)} (\k - \k')$.  Taking $H$, $\lambda$ and $\mu$ to be constant, the equations of motion following from \eqref{action} can be written as
\begin{align}
\!\!\dv{z} \left( \dv{z}  \varphi  + \frac{\lambda}{z} \sigma \right)  - \frac{2}{z}\left( \dv{z}  \varphi  + \frac{\lambda}{z} \sigma \right) +  \varphi  =0  , \label{eom-1} \\
\!\!\dv[2]{z} \sigma - \frac{2}{z} \dv{z} \sigma + \left[1 +  \frac{\mu^2}{H^2 z^2} \right] \sigma  = \frac{\lambda}{z} \left( \dv{z}  \varphi  + \frac{\lambda}{z} \sigma \right), \label{eom-2} 
\end{align}
where $z \equiv  k / H a(t)$.  The variable $z$ measures the physical wavenumber $k/a(t)$ in units of the Hubble scale $H$. At early times, $z\gg1$ and the mode is deep inside the horizon; horizon crossing occurs at $z=1$; and at late times, $z\to0$ and the mode becomes superhorizon.

%%%%%%%%%%%%%%%%%%%%%%%%%%%%%%%%%%%%%%%%%%%%%%%%%%%%%%%%%%%%%%%%%%%%%%%%%%%%%%%%%%%%%%%%%%%%%%%%%%%%%%%%%%%%%%%%%%%%%%%%%%%%%%%%%%%%%%%%%%%%%%%%%%%%%%%%%%%%%%%%%%%%%%%%%%%%%%%%%%%%%%%%%%%%%%%%%%%%%%%%%%%%%%%%%%%%%%%%%%%%%%%%%%%%%%%%%%%%%%%%%%%%%%%%%%%%%%%%%%%%%%%%%%%%%%

 \noindent \emph{\bf Bogoliubov Decomposition.--}
In the absence of interactions ($\lambda = 0$), the perturbations $\varphi$ and $\sigma$ decouple. The corresponding mode functions are
\be
\varphi_b (k, z) = \delta_{b1} u_0(k,z) ,  \quad 
\sigma_b (k, z) =  \delta_{b2} u_\mu (k,z) , 
\ee
where $u_0 (k,z)$ and $u_\mu (k,z)$ are the standard Bunch--Davies mode functions for a massless and a massive scalar, respectively. They are given by
\begin{align}
u_0 (k,z) &=   
\frac{iH}{\sqrt{2 k^3}} 
\left(1 - i z \right) e^{ i z } , \label{def-u-0}
\\
u_\mu (k,z) 
&= 
- \frac{H}{\sqrt{2 k^3}}  
\sqrt{\frac{ \pi}{2 }} 
e^{i\theta_\nu} 
z^{ 3/2} 
H_\nu^{(1)} \left( z \right) , \label{def-u-mu}
\end{align}
where $H_\nu^{(1)}(z)$ is the Hankel function of the first kind, with index $\nu = \sqrt{\frac{9}{4} - \frac{\mu^2}{H^2}}$, and $\theta_\nu = \frac{\pi}{2}\left(\nu-\frac32\right)$. Using these decoupled solutions as a basis, we write the interacting mode functions as~\cite{Parra:2024usv}
\begin{align}
\varphi_b (k, z) &= f_{b} (z) u_0(k,z) + \bar f_{b}(z) u^*_0(k,z) , \label{varphi-u} \\
\sigma_b (k, z) &= s_{b} (z) u_\mu (k,z) + \bar s_{b} (z) u^*_\mu (k,z) . \label{sigma-u}
\end{align}
The functions $f_{b} (z)$, $\bar f_{b} (z)$, $s_{b} (z)$, and $\bar s_{b} (z)$ are time-dependent Bogoliubov coefficients (here the bar distinguishes between different coefficients while complex conjugation is denoted with a star).  

The equations of motion \eqref{eom-1} and \eqref{eom-2} can be recast as first-order equations for the Bogoliubov coefficients. Ordering them as $(f_b, \bar f_b , s_b , \bar s_b)^T$, one obtains~\cite{Fumagalli:2021mpc, Parra:2024usv}
\be\label{first-order-eq-bogos}
 z \dv{z} \left(\begin{array}{c} f_b \\ \bar f_b \\ s_b \\ \bar s_b 
\end{array}\right)  =  \left(\begin{array}{cc} 0 & \mathcal P (z ) \\ 
\mathcal Q (z ) & 0 
\end{array}\right)  \left(\begin{array}{c} f_b \\ \bar f_b \\ s_b \\ \bar s_b 
\end{array}\right) ,
\ee
where $\mathcal P (z)$ and $\mathcal Q (z)$ are the $2 \times 2$ matrices
\begin{align}
\mathcal P (z) &=
- i \frac{\lambda k^3}{H^2 z^{2}}
\left(\begin{array}{cc}
{u_0^{*}}' u_\mu & {u_0^{*}}' u_\mu^{*} \\
- u_0' u_\mu & - u_0' u_\mu^*
\end{array}\right) ,
\\
\mathcal Q (z) &=
- i \frac{\lambda k^3}{H^2 z^{2}}
\left(\begin{array}{cc}
u_0' u_\mu^* & {u_0^{*}}' u_\mu^* \\
- u_0' u_\mu & - {u_0^{*}}' u_\mu
\end{array}\right) .
\end{align}
In these expressions, primes denote derivatives with respect to $z$. To solve \eqref{first-order-eq-bogos}, we impose initial conditions at a fiducial time $z=z_0$ deep inside the horizon, with $z_0 \gg 1$. We choose
\begin{align} 
( f_1 , \bar f_1 , s_1 , \bar s_1 )(z_0)  &= (1 , 0 , 0 , 0) , \label{init-1} \\
( f_2 , \bar f_2 , s_2 , \bar s_2 )(z_0)  &= (0 , 0 , 1 , 0)  . \label{init-2}
\end{align}
In the absence of interactions, these relations reduce to the standard Bunch--Davies initial conditions.

%%%%%%%%%%%%%%%%%%%%%%%%%%%%%%%%%%%%%%%%%%%%%%%%%%%%%%%%%%%%%%%%%%%%%%%%%%%%%%%%%%%%%%%%%%%%%%%%%%%%%%%%%%%%%%%%%%%%%%%%%%%%%%%%%%%%%%%%%%%%%%%%%%%%%%%%%%%%%%%%%%%%%%%%%%%%%%%%%%%%%%%%%%%%%%%%%%%%%%%%%%%%%%%%%%%%%%%%%%%%%%%%%%%%%%%%%%%%%%%%%%%%%%%%%%%%%%%%%%%%%%%%%%%%%%

 \noindent \emph{\bf Exact solutions.--}
We now turn to the exact solution of the coupled system. Starting from \eqref{first-order-eq-bogos}, and assuming $\lambda \neq 0$, one may eliminate three of the four Bogoliubov coefficients to obtain a fourth-order differential equation for any one of them. In particular, $f= f(z)$ satisfies
\vspace{-1pt}
\begin{align}
 \bigg[ &
z^2\dv[4]{z}
+4z(1+iz)\dv[3]{z}
+\Big(\frac{9}{4}-\nu^2+10iz-4z^2 \Big)  \dv[2]{z}  \nn \\
& +\Big(i \frac{5}{2}-2 i \nu^2 -4z \Big)  \dv{z}
-\lambda^2
\bigg]  f = 0 . 
\label{4th-ode-general-nu}
\end{align}
The remaining coefficients, $\bar f(z)$, $s(z)$, and $\bar s(z)$, are reconstructed from $f (z)$ with the help of \eqref{first-order-eq-bogos}. In particular $\bar f$ can be obtained from $f$ as:
\begin{align}
\bar f
= & 
e^{2iz}
\bigg[
1
-\frac{2i}{\lambda^2}
\bigg(
z^2\dv[3]{z}
+2z(1+iz)\dv[2]{z} \nn\\ 
&\qquad +\Big(
\frac14-\nu^2+2iz
\Big)\dv{z}
\bigg)
\bigg] f .
\label{barf-f}
\end{align}
Equation~\eqref{4th-ode-general-nu} provides an equivalent fourth-order formulation of the original first-order system~\eqref{first-order-eq-bogos}. Its four independent solutions can be written as
\begin{align}
f_{\rm I}^{(w)}(z)
&=
\frac{e^{\pi w\lambda/4}}{\sqrt{2}}\,
\widehat{\mathcal D}_\nu^{(w)}
U\left(
\frac{iw\lambda}{2},1,-2iz
\right),
\label{f-sol-1}
\\
 f_{\rm II}^{(w)} (z) & =\bigl(\bar f_{\rm I}^{(w)} (z)\bigr)^* ,
\label{f-sol-2}
\end{align}
where $w = \pm 1$ labels the two branches related by $\lambda \to - \lambda$. Note that $f_{\rm II}^{(w)}$ is obtained from $f_{\rm I}^{(w)}$ by applying \eqref{barf-f} on it  and taking the complex conjugate~\cite{note-solution}. In the previous solutions, $U(a,b,z)$ is Tricomi's confluent hypergeometric function, which admits the following integral representation
\be
U(a,b,x)=\frac{1}{\Gamma(a)}
\int_0^\infty \! \dd{t}\,
e^{-xt}t^{a-1}(1+t)^{b-a-1},
\label{Tricomi-integral}
\ee
initially valid for $\mathrm{Re}\,a>0$ and
$\mathrm{Re}\,x>0$, and extended to the parameters
in \eqref{f-sol-1} by analytic continuation, with
$x=-2iz$ approached from $\mathrm{Re}\,x>0$.  Finally, the operator $\mathcal{\widehat{D}}_{\nu}^{(w)}$ is defined as
\be
\mathcal{\widehat{D}}_{\nu}^{(w)}
\equiv
{}_2F_1\left(
\frac{1}{2} - \nu , \frac{1}{2} + \nu;
1 + i w \lambda;
\frac{i}{2}\dv{z}
\right),
\ee
where ${}_2F_1$ is the Gauss hypergeometric function. In this way, \eqref{f-sol-1} and $\bar f_{\rm I}^{(w)} (z)$  entering \eqref{f-sol-2}, may be written as~\cite{note-integral} 
\begin{align}
f_{\rm I}^{(w)}(z) \!
&= \! 
\frac{e^{\frac{\pi w\lambda}{4}}}
{\sqrt{2}\,\Gamma\left(\frac{iw\lambda}{2}\right)} \!\!
\int_0^\infty  \!\!\!\! \dd{t} \frac{e^{2izt} }{t} F_w(t) \! 
\left(\frac{t}{1+t}\right)^{\frac{iw\lambda}{2}}  \!\! \!\!  \!\! \!\!  
 , \label{f-sol-1-alt} \\
\bar f_{\rm I}^{(w)}(z) \!
&= \! 
\frac{e^{2 i z + \frac{\pi w\lambda}{4} } }
{\sqrt{2}\,\Gamma\left(\frac{iw\lambda}{2}\right)} \!\!
\int_0^\infty  \!\!\!\! \dd{t} \frac{e^{2izt} }{t} F_w(t) \!
\left(\frac{t}{1+t}\right)^{1+ \frac{ iw\lambda}{2}}  \!\! \!\!  \!\! \!\! \!\! \!\!  , \quad \label{f-sol-2-alt}
\end{align}
where $F_w(t)$ is defined as
\be
F_w(t) \equiv   {}_2F_1\left(
\frac12-\nu,\frac12+\nu;1+iw\lambda;-t
\right) . \label{def-F_w}
\ee
Notice that for $\nu = 1/2$ (corresponding to a conformally coupled massless scalar in de Sitter space) the differential operator reduces to the identity: $\mathcal{\widehat{D}}_{\nu}^{(w)}=1$. Finally, the general solution for each mode $b$ is the linear combination
\be
f_b(z) =\sum_{w = \pm 1} A_b^{(w)} f_{\rm I}^{(w)}(z) + \sum_{w = \pm 1} B_b^{(w)} f_{\rm II}^{(w)}(z), \label{f-general-sol}
\ee
where $A_b^{(w)}$ and $B_b^{(w)}$ are integration constants restricted to satisfy $\sum_b
 A_b^{(w)}A_b^{(w')*} - \sum_b B_b^{(w)*}B_b^{(w')}  = \delta_{ww'}$ and $\sum_b  A_b^{(w)}B_b^{(w')*} - \sum_b B_b^{(w)*}A_b^{(w')}  = 0$. 

 \vspace{6pt}

%%%%%%%%%%%%%%%%%%%%%%%%%%%%%%%%%%%%%%%%%%%%%%%%%%%%%%%%%%%%%%%%%%%%%%%%%%%%%%%%%%%%%%%%%%%%%%%%%%%%%%%%%%%%%%%%%%%%%%%%%%%%%%%%%%%%%%%%%%%%%%%%%%%%%%%%%%%%%%%%%%%%%%%%%%%%%%%%%%%%%%%%%%%%%%%%%%%%%%%%%%%%%%%%%%%%%%%%%%%%%%%%%%%%%%%%%%%%%%%%%%%%%%%%%%%%%%%%%%%%%%%%%%%%%%

 \noindent \emph{\bf Initial conditions.--}
Let us fix the integration constants by imposing (\ref{init-1}) and (\ref{init-2}). To proceed, notice that at large values of $z$, the behavior of \eqref{f-sol-1-alt} is governed by the lower limit $t=0$, where both $(1+t)^{-iw\lambda/2}$ and the hypergeometric function become $1+\mathcal O(t)$. Keeping their leading terms leaves the elementary Laplace integral $\Gamma(iw\lambda/2)(-2iz)^{-iw\lambda/2}$, and so one recovers $f_{\rm I}^{(w)}(z) \simeq\frac{1}{\sqrt{2}}
e^{-\frac{iw\lambda}{2}\ln(2z) } $, with corrections suppressed by powers of $z^{-1}$. Similarly, from \eqref{f-sol-2-alt} it follows that $f_{\rm II}^{(w)}(z) \simeq - \frac{w \lambda}{4\sqrt{2} z}
e^{-2 i z + \frac{iw\lambda}{2}\ln(2z) }$. Inserting these expressions in \eqref{f-general-sol} and using~\eqref{first-order-eq-bogos} gives the corresponding asymptotic behavior for $\bar f$, $s$, and $\bar s$. Then, the initial conditions \eqref{init-1} and \eqref{init-2} give the integration constants as
\be
A_1^{(w)}  = i w A_2^{(w)} =
\frac{1}{\sqrt{2}} e^{
\frac{iw\lambda}{2}\ln(2z_0) }
+
\mathcal O(z_0^{-1}),
\label{init-A1}
\ee
while $B_b^{(w)}$ are all of order $\mathcal O(z_0^{-1})$. Therefore, in the Bunch--Davies limit $z_0\to+\infty$, the physical solutions receive contributions only from the two functions $f_{\rm I}^{(w)}(z)$.

%%%%%%%%%%%%%%%%%%%%%%%%%%%%%%%%%%%%%%%%%%%%%%%%%%%%%%%%%%%%%%%%%%%%%%%%%%%%%%%%%%%%%%%%%%%%%%%%%%%%%%%%%%%%%%%%%%%%%%%%%%%%%%%%%%%%%%%%%%%%%%%%%%%%%%%%%%%%%%%%%%%%%%%%%%%%%%%%%%%%%%%%%%%%%%%%%%%%%%%%%%%%%%%%%%%%%%%%%%%%%%%%%%%%%%%%%%%%%%%%%%%%%%%%%%%%%%%%%%%%%%%%%%%%%%

\begin{figure}[t]
\centering
\includegraphics[width=\columnwidth]{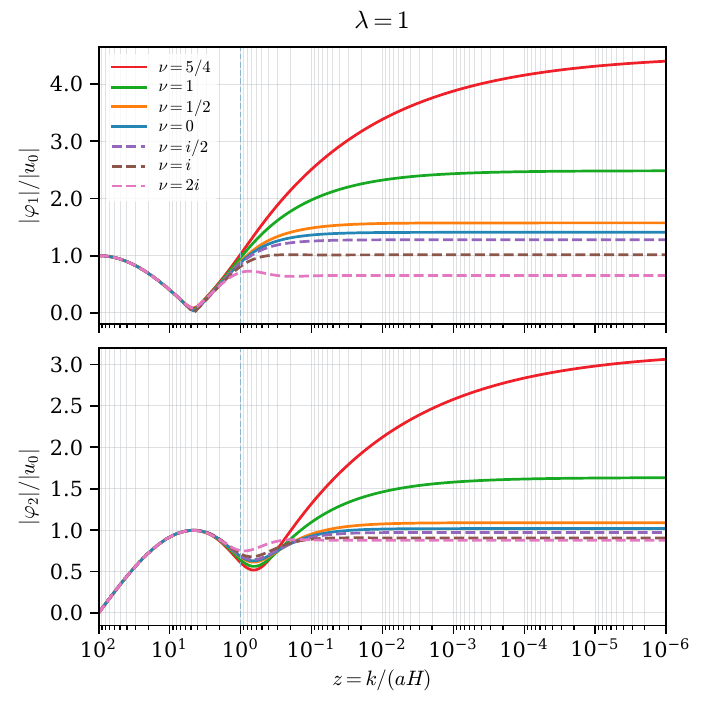}
\caption{
Normalized mode amplitudes $|\varphi_b|/|u_0|=|(1-iz)e^{iz}f_b-(1+iz)e^{-iz}\bar f_b|/\sqrt{1+z^2}$ for $b=1$ (top) and $b=2$ (bottom), with $\lambda = 1$ and selected values of $\nu$. Initial conditions~(\ref{init-1}) and~(\ref{init-2}) are imposed at $z_0=100$. The vertical dashed line marks horizon crossing, $z = 1$.
}
\label{fig:solutions}
\vspace{-15pt}
\end{figure}

\noindent \emph{\bf Primordial power spectrum.--}
The main observable of interest is the dimensionless primordial power spectrum $\Delta_\zeta(k,z)$ evaluated at the end of inflation ($z\ll1$) which is $\Delta_{\zeta} (k , z) = \frac{k^3}{4 \pi^2 \epsilon} \sum_b |\varphi_b (z)|^2$. Fig.~\ref{fig:solutions} shows the normalized mode amplitudes $|\varphi_b|/|u_0|$ for $\lambda=1$ and selected values of $\nu$.  
Using \eqref{varphi-u}  and \eqref{sigma-u}, on superhorizon scales, $\Delta_{\zeta}(k,z)$ becomes
\be
\Delta_{\zeta}(k,z) = \Delta_0(k)
\sum_b\left|f_b(z)-\bar f_b(z)\right|^2,
\label{power-bogos}
\ee
where $\Delta_0(k)=H^2/(8\pi^2\epsilon)$ is the single-field result. Using \eqref{f-sol-1-alt}  and \eqref{f-sol-2-alt} to compute the late time difference gives 
\begin{align}
 f_{\rm I}^{(w)}-\bar f_{\rm I}^{(w)} 
& = - 2iz f_{\rm I}^{(w)} (z) \nn \\
 & \hspace{-45pt} +  \frac{e^{\pi w\lambda/4}}{\sqrt2 \Gamma(iw\lambda/2)} \!  \int_0^\infty \!\! \!\!  \dd t\,
\frac{e^{2 i z t} t^{\frac{ iw\lambda}{2}-1}}{(1+t)^{1 + \frac{ iw\lambda}{2}}}F_w(t) + \mathcal O (z). \label{diff-f}
\end{align}
To evaluate this expression we must consider the massive case ($\mu^2>0$) separately from the massless case ($\nu=3/2$), as in the latter $\sigma$ does not decay as $z \to 0$~\cite{Achucarro:2019lgo}, sourcing the logarithmic evolution of $\zeta$ on superhorizon scales. 

\begin{figure}[b]
\centering
\includegraphics[width=0.9\columnwidth]{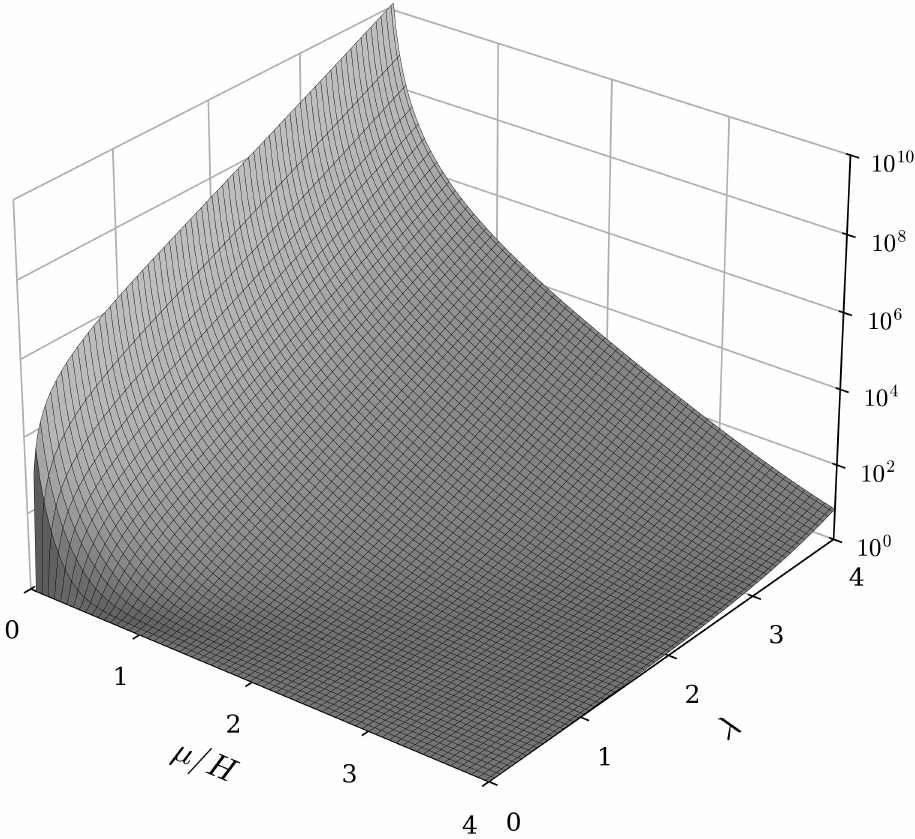}
\caption{$\Delta_\zeta/\Delta_0$ given in Eq.~\eqref{eq:power-spectrum-general-nu} as a function of $\mu/H$ and $\lambda$. }
\label{fig:power}
\end{figure}

Let us start with the massive case. Taking the limit $z\to0$ in \eqref{diff-f}, the first term vanishes, and we may set $z=0$ in the integral. To solve the integral of the second line we set $x=t/(1+t)$, and use Pfaff's transformation followed by Euler's integral. The result is
\begin{align}
f_{\rm I}^{(w)}-\bar f_{\rm I}^{(w)} 
& =\frac{e^{\pi w\lambda/4}}{\sqrt2} \frac{\Gamma\left(\frac32-\nu\right)}
{\Gamma\left(\frac32-\nu+\frac{iw\lambda}{2}\right)}\nn\\
& \quad \times {}_3F_2\!\left(\begin{matrix}
\frac12-\nu,\frac12-\nu+iw\lambda,\frac{iw\lambda}{2}\\
1+iw\lambda,\frac32-\nu+\frac{iw\lambda}{2}
\end{matrix};1\right) .
\end{align}
Applying Dixon's summation formula allows one to express ${}_3F_2$ in terms of gamma functions. Summing the two modes with the initial coefficients of Eq.~\eqref{init-A1} cancels the interference terms and the $z_0$-dependent phases, giving
\be
\frac{\Delta_\zeta}{\Delta_0}
=
\left|
\frac{
\Gamma\left(\frac34-\frac{\nu}{2}\right)
\Gamma\left(\frac34+\frac{\nu}{2}\right)
}{
\Gamma\left(\frac34-\frac{\nu}{2}+\frac{i\lambda}{2}\right)
\Gamma\left(\frac34+\frac{\nu}{2}+\frac{i\lambda}{2}\right)
}
\right|^2 .
\label{eq:power-spectrum-general-nu}
\ee
This expression reduces to $\Delta_\zeta/\Delta_0=1$ when $\lambda=0$. A simple illustrative case is $\nu=1/2$, corresponding to $\mu^2/H^2=2$, or equivalently to a conformally coupled massless scalar in de Sitter space. In this case, $\mathcal{\widehat D}_{\nu}^{(w)}=1$, and the result simplifies to 
\be
\frac{\Delta_\zeta}{\Delta_0}
=
\frac{\sinh(\pi\lambda)}{\pi\lambda} .
\ee
The heavy-field regime, $\mu^2/H^2>9/4$, follows by analytic continuation, $\nu\to i\rho$, with $\rho=\sqrt{\mu^2/H^2- 9/4}$. After simplification, the power spectrum becomes
\be
\frac{\Delta_\zeta}{\Delta_0}
=
\frac{
\left|
\Gamma\left(\frac34+\frac{i\rho}{2}\right)
\right|^4
}{
\left|
\Gamma\left(\frac34+\frac{i}{2}(\lambda+\rho)\right)
\right|^2
\left|
\Gamma\left(\frac34+\frac{i}{2}(\lambda-\rho)\right)
\right|^2
} .
\ee
In the large-mass limit, $\mu^2\gg H^2$ (or $\rho\gg1$), this expression admits a simple expansion. At fixed $\lambda$, one finds:
\be
\frac{\Delta_\zeta}{\Delta_0} =
1+\frac{\lambda^2}{2\rho^2}
+
\frac{3\lambda^2(1+\lambda^2)}{8\rho^4}
+\mathcal O(\rho^{-6}) .
\ee
The leading correction reproduces the standard effective-single-field result obtained by integrating out a heavy isocurvature mode~\cite{Tolley:2009fg, Achucarro:2010da, Cespedes:2012hu, Achucarro:2012sm, Chen:2012ge, Achucarro:2012yr, Noumi:2012vr, Castillo:2013sfa,Gwyn:2012mw}, while the subleading terms give corrections to the heavy-mass expansion. This provides a nontrivial check of the expression above. Fig.~\ref{fig:power} shows the power-spectrum ratio $\Delta_\zeta/\Delta_0$ as a function of $\mu/H$ and $\lambda$, in the limit $z\to 0$. As expected, on superhorizon scales, the amplitude of $\zeta$ is enhanced substantially for small values of $\mu /H$ and large values of $\lambda$.

Now for the massless limit. At $\nu=3/2$, \eqref{def-F_w} reduces to $F_w(t)=1+2t/(1+iw\lambda)$. The factor multiplying $e^{2izt}$ in \eqref{f-sol-1-alt} therefore approaches $2/(1+iw\lambda)$ at large $t$, giving the $1/z$ growth of $f_{\rm I}^{(w)}$. With the contour rotation of footnote~\cite{note-integral}, this yields $-2izf_{\rm I}^{(w)} \to \sqrt2 \,e^{\pi w\lambda/4}/[\Gamma(iw\lambda/2)(1+iw\lambda)]$. The integral in the second line of \eqref{diff-f} separates into a finite contribution, tending to $2/(iw\lambda)$, and a logarithmic contribution. The small-argument expansion of $U(1+iw\lambda/2,1,-2iz)$ gives the latter as $-2[\ln(-2iz)+\psi(1+iw\lambda/2)+2\gamma_{\rm E}]/(1+iw\lambda)$, up to terms vanishing as $z\to0$. Here $\psi(x)=\Gamma'(x)/\Gamma(x)$ is the digamma function and $\gamma_{\rm E}$ is Euler's constant. Combining these contributions using $\ln(-2iz)=\ln(2z)-i\pi/2$ for $z>0$ and $\psi(1+x)=\psi(x)+1/x$, we obtain:
\begin{align}
f_{\rm I}^{(w)}-\bar f_{\rm I}^{(w)}
&=
\frac{e^{\pi w\lambda/4}}{
\sqrt2\,\Gamma\left(\frac{i w \lambda}{2}\right)(1+ i w \lambda)
}  \bigg[4 + \frac{2 i w}{\lambda} \nn \\
&\hspace{-30pt}
-2\left(
\ln(2z)-\frac{i\pi}{2}
+\psi\left(\frac{i w \lambda}{2}\right)
+2\gamma_{\rm E}
\right)
\bigg] .
\end{align}
Using the initial coefficients of \eqref{init-A1} in the two-mode sum \eqref{power-bogos}, the $z_0$-dependent phases cancel, yielding the result
\begin{align}
\frac{\Delta_\zeta}{\Delta_0}
&=
\frac{\lambda\sinh(\pi\lambda)}
{\pi(1+\lambda^2)} \!
\left[
2 - \ln(2 z_{\rm end})-2\gamma_{\rm E}
-\mathrm{Re}\,\psi\left(\frac{i\lambda}{2}\right)
\right]^2 \nn \\
&\quad
+
\frac{\pi\lambda}{2(1+\lambda^2)}
\coth\left(\frac{\pi\lambda}{2}\right) ,
\end{align}
where $z_{\rm end}$ is the value of $z$ at the end of inflation. This expression displays the secular superhorizon growth produced by the mixing $\lambda$~\cite{Achucarro:2019lgo}. For $\lambda\ll1$ it reduces to $\Delta_\zeta/\Delta_0=1+\lambda^2\big[(2-\gamma_{\rm E}-\ln 2z_{\rm end})^2+\pi^2/12-1\big]+\mathcal O(\lambda^4)$, in agreement with the perturbative result of Ref.~\cite{Parra:2024usv}.

%%%%%%%%%%%%%%%%%%%%%%%%%%%%%%%%%%%%%%%%%%%%%%%%%%%%%%%%%%%%%%%%%%%%%%%%%%%%%%%%%%%%%%%%%%%%%%%%%%%%%%%%%%%%%%%%%%%%%%%%%%%%%%%%%%%%%%%%%%%%%%%%%%%%%%%%%%%%%%%%%%%%%%%%%%%%%%%%%%%%%%%%%%%%%%%%%%%%%%%%%%%%%%%%%%%%%%%%%%%%%%%%%%%%%%%%%%%%%%%%%%%%%%%%%%%%%%%%%%%%%%%%%%%%%%

\noindent \emph{\bf Discussion.--}
Equations~\eqref{f-sol-1} and \eqref{f-sol-2} are the central result of this Letter. They provide exact analytic control over the linear dynamics of coupled curvature and isocurvature perturbations in a two-field, constant-turn inflationary background, for arbitrary values of the interaction strength $\lambda$ and entropy mass $\mu$. This includes the strongly mixed light-field regime associated with rapid-turn inflation, which has often required numerical treatments or analytic approximations of special limits. As an application, we used these solutions to derive the amplitude of the primordial power spectrum $\Delta_{\zeta}$, obtaining expressions that interpolate between the light- and heavy-field regimes and between weak and strong coupling.

The strongly mixed regime is of direct relevance to supergravity and string-theory constructions, where curved field spaces naturally sustain strongly non-geodesic trajectories, allowing inflation to proceed on potentials too steep for single-field slow-roll~\cite{Brown:2017osf, Mizuno:2017idt, GarciaSaenz:2018ifx, Aragam:2019omo, Bjorkmo:2019fls, Aragam:2020uqi}. In such models the turning rate is typically large and the entropic mode is light or even tachyonic in terms of $m^2=\mu^2-\lambda^2H^2$~\cite{Aragam:2021scu}, precisely where perturbative treatments break down and effective single-field descriptions have limited validity, while our solutions remain valid.

These results also provide a new analytic starting point for the phenomenology of multifield inflation, particularly in regimes relevant to the cosmological-collider program~\cite{Arkani-Hamed:2015bza, Lee:2016vti, Flauger:2016idt, Chen:2016uwp, Xianyu:2025lbk, Kumar:2026ogn, Kumar:2026dih}. Because the solutions~\eqref{f-sol-1} and \eqref{f-sol-2} are nonperturbative in the mixing strength, they provide controlled access to strongly mixed regimes in which an expansion in $\lambda$ is unreliable. For instance, finite-duration turns could be modeled by matching constant-turn solutions across successive time intervals. More broadly, these solutions should provide useful building blocks for analytic studies of higher-point functions, particle production, primordial-black-hole formation, scalar-induced gravitational waves, and loop corrections to primordial correlators~\cite{Weinberg:2005vy, Weinberg:2006ac, Senatore:2009cf, WangXianyuZhong:2021, XianyuZhang:2022}. A detailed computation of the bispectrum and an analysis of its squeezed limit are presented in the companion article~\cite{Huenupi:2026aqc}.

Several extensions deserve further attention. Departures from exact de Sitter and from constant $\lambda$ and $\mu$, of order $\epsilon$, $\dot\epsilon/(H\epsilon)$, $\dot\lambda/(H\lambda)$ and $\dot\mu/(H\mu)$, can be included perturbatively and break the scale invariance of the power spectrum. Since $\Delta_\zeta$ is fixed by observations, a large enhancement necessarily lowers the tensor-to-scalar ratio, $r=16\epsilon\,\Delta_0/\Delta_\zeta$; in that regime, the validity of the linear treatment rests on the size of the cubic couplings. Finally, models with more than two fields involve additional entropy directions. We leave these questions, as well as the remaining applications, for future work.

%%%%%%%%%%%%%%%%%%%%%%%%%%%%%%%%%%%%%%%%%%%%%%%%%%%%%%%%%%%%%%%%%%%%%%%%%%%%%%%%%%%%%%%%%%%%%%%%%%%%%%%%%%%%%%%%%%%%%%%%%%%%%%%%%%%%%%%%%%%%%%%%%%%%%%%%%%%%%%%%%%%%%%%%%%%%%%%%%%%%%%%%%%%%%%%%%%%%%%%%%%%%%%%%%%%%%%%%%%%%%%%%%%%%%%%%%%%%%%%%%%%%%%%%%%%%%%%%%%%%%%%%%%%%%%

\begin{acknowledgments}

\vspace{2pt}
\noindent\emph{\bf Acknowledgments.--} We wish to thank Ana Ach\'ucarro, Perseas Christodoulidis, Gabriel Mar\'in Mac\^edo, Hayden Lee, Anish Pandya, Nicol\'as Parra, S\'ebastien Renaux-Petel, and Crist\'obal Zenteno for useful discussions and comments. CM acknowledges support from FONDECYT grant 1231250, from the Centro de Modelamiento Matemático (CMM) BASAL fund FB210005, and from the PRISMALab Group at CMM. GAP acknowledges support from ANID through FONDECYT Regular grants 1210876 and 1251511. SS acknowledges support from FAPEI through grant FP2688125 and from DICREA UBB through grant IN2686604.
\end{acknowledgments}


\begin{thebibliography}{199}
\bibitem{note-mass}
Note that expanding the first term in \eqref{action} shifts the coefficient of $-\tfrac12\sigma^2$ from $\mu^2$ to $m^2=\mu^2-\lambda^2H^2$. We find it more natural to work with $\mu$, the mass appearing in the Hamiltonian, in which the mixing enters only as $\lambda H\,\Pi_\varphi\sigma$, with $\Pi_\varphi$ the momentum conjugate to $\varphi$.
\bibitem{note-solution}
For real $\lambda$ and $\mu^2$, Eqs.~\eqref{eom-1} and \eqref{eom-2} are real, so the complex conjugate of any solution $(\varphi_b,\sigma_b)$ is a solution. By \eqref{varphi-u} and \eqref{sigma-u}, conjugation maps $(f,\bar f,s,\bar s)\to(\bar f^*,f^*,\bar s^*,s^*)$. Hence, if $f$ solves \eqref{4th-ode-general-nu}, then $\bar f^*$ also does.
\bibitem{note-integral}
For real $\lambda$ the integrand of \eqref{f-sol-1-alt} behaves as $t^{-1+iw\lambda/2}$ near $t=0$, so the analytic continuation from ${\rm Re}\,a>0$ must be made explicit. Integrating by parts once and rotating the contour to $t=i\tau$ gives, for $z>0$,
$f_{\rm I}^{(w)}(z) = -\frac{i}{\sqrt{2}\,\Gamma(1+\frac{iw\lambda}{2})}\int_0^\infty \dd{\tau}\, \tau^{\frac{iw\lambda}{2}}\, g_w'(i\tau)$, with $g_w(t)\equiv e^{2izt}(1+t)^{-\frac{iw\lambda}{2}}F_w(t)$ and $F_w'(t) = \frac{\nu^2-1/4}{1+iw\lambda}\,{}_2F_1(\frac32-\nu,\frac32+\nu;2+iw\lambda;-t)$.  Similarly,
$\bar f_{\rm I}^{(w)}(z) = \frac{i\,e^{2iz}}{\sqrt{2}\,\Gamma(\frac{iw\lambda}{2})}\int_0^\infty \dd{\tau}\, \tau^{\frac{iw\lambda}{2}}(1+i\tau)^{-1-\frac{iw\lambda}{2}}\, e^{-2z\tau} F_w(i\tau)$.
Both integrals converge absolutely and exponentially, and are straightforward to evaluate numerically. The rotation is allowed because the integrands are analytic in the first quadrant, with branch points only at $t=0$ and $t\le-1$. The phase $(i\tau)^{iw\lambda/2} =e^{-\pi w\lambda/4}\tau^{iw\lambda/2}$ 
cancels the prefactor $e^{\pi w\lambda/4}$.
\end{thebibliography}
\end{document}